\documentstyle[12pt]{article}
\newcommand{\beq}{\begin{equation}}
\newcommand{\eeq}{\end{equation}}
\begin{document}

\begin{titlepage}
\begin{center}
{\hbox to\hsize{ \hfill IZTECH-P04/2004}}

\bigskip
\vspace{3\baselineskip}

{\Large \bf  Effects of Flavor Violation on Split Supersymmetry}

\bigskip

\bigskip

{\bf  Durmu{\c s} A. Demir}\\
\smallskip
{\small \it Department of Physics, Izmir Institute of Technology,
Izmir, TR35430, Turkey}

\bigskip

\vspace*{.5cm}

{\bf Abstract}\\
\end{center}
\noindent 
We discuss consequences of flavor mixings in 
the scalar fermion sector of supersymmetric 
models endowed with ultra heavy scalars. We 
find that, under extreme fine-tunings different
than but similar in size to that needed to
obtain a light Higgs doublet, intergenerational 
mixings generically lead to light sfermion 
states which facilitate a number of phenomena 
ranging from rare processes to electric dipole moments. 
The number of scalar fermions to be discovered 
by collider searches is fewer than those in
a complete supersymmetric model.

\bigskip

\bigskip

\end{titlepage}

The models of physical phenomena are generically based on the
principle of naturalness, that is, the mother Nature strongly
disfavors fine-tunings. The present understanding of the four
fundamental forces of Nature faces with two naturalness problems:
($i$) Higgs boson mass exhibits a quadratic sensitivity to the
ultraviolet (UV) cut-off of the standard electroweak theory (SM),
and ($ii$) experimental value of the cosmological constant (CC)
turns out to be far below the theoretical estimates. Concerning
the former, a resolution lies in embedding the SM into a UV-safe
extension above Fermi energies. A solution for the CC problem, on
the other hand, might eventually require modifications in the
gravitational dynamics of the vacuum energy. If Nature indeed
prefers a supersymmetric organizing principle above the Fermi
scale then Higgs boson mass gets stabilized against unnatural
quantum fluctuations. Furthermore, supersymmetric nature of
interactions ($i$) predicts the unification of three gauge forces
near the scale of strong gravitational interactions, ($ii$) realizes
radiative electroweak breaking, and ($iii$) provides a viable dark
matter candidate. In spite of these rather appealings aspects,
however, supersymmetry does not offer a solution to the CC problem
as it cannot be a good symmetry of Nature below the electroweak
scale. In fact, a solution to the CC problem requires 'new
physics' to show up at a scale around the neutrino mass. Given
these features of the two naturalness problems, a simultaneous and
unified resolution, as recently proposed by Arkani-Hamed and
Dimopoulos \cite{ad}, might be that Nature is inherently
fine-tuned.  In other words, Nature might admit small numbers
$\epsilon_{CC}$ and $\epsilon_{h}$ such that
\begin{eqnarray}
\label{fine} \Lambda \sim \epsilon_{CC}\, M_{SUSY}^4\;\;,\;\;\;
m_h^2 \sim \epsilon_h\, M_{SUSY}^2
\end{eqnarray}
where the scale of supersymmetry breaking, $M_{SUSY}$, can be well 
inside the Planckian territory or in some intermediate domain
depending on what mechanism is responsible for the breaking. The
smallness of $\epsilon_{CC,h}$ is a measure of the UV-sensitivity
of the quantity under concern: for $M_{SUSY}\sim M_{Pl}$ one has
$\epsilon_{CC}\sim 10^{-120}$ and $\epsilon_{h} \sim 10^{-32}$. In
this picture it is the presence of fine-tunings, rather than the
renormalization group flow of masses, that provides a SM-like
Higgs doublet to condense for breaking the gauge symmetry. Highly
appealing aspects of the supersymmetric models, unification of the
gauge couplings and presence of a dark matter candidate, are still
maintained thanks to the chiral protection of gaugino and Higgsino
masses \cite{dimopoulos,gm}. In the minimal supersymmetric model (MSSM), for instance,
the $\mu$ parameter, gaugino masses and triscalar couplings all
explicitly break the continuous $R$ invariance of the model
(because of which they are endowed with CP--odd phases).

In the following we will focus on the scalar fermion sector of the
MSSM, endowed with mass hierarchies pertinent to split supersymmetry \cite{ad,followups},
and show that part of the sfermions necessarily weigh at the weak
scale if ($i$) the sfermion mass-squared matrices exhibit flavor mixings
and ($ii$) extreme fine-tunings (different than but similar in size
to that needed to induce a light Higgs doublet) are allowed.
The motivations for such an analysis are twofold: ($i$) the 
mechanism that breaks the supersymmetry does not need 
to be flavor-blind, and ($ii$) there are no chiral symmetries 
that can protect any entry of a generic
sfermion mass matrix. The detailed discussions below will give
rise to the conclusion that, non-observation of light sfermion
states at the LHC will guarantee approximate flavor-blindness of
supersymmetry breaking.

It is useful to start our analysis with a brief review of the main
points made in \cite{ad}. The superpotential of the MSSM
\begin{eqnarray}
\label{superpot} \widehat{W} = \mu \widehat{H}_u\cdot
\widehat{H}_d + \widehat{Q} \cdot \widehat{H}_u {\bf Y_u}
\widehat{U}^c\: +\widehat{H}_d \cdot \widehat{Q} {\bf Y_d}
\widehat{D}^c\: + \widehat{H}_d \cdot \widehat{L} {\bf Y_e}
\widehat{E}^c
\end{eqnarray}
encodes the rigid parameters of the model: the Higgsino mass $\mu$
and the Yukawa matrices ${\bf Y_{u,d,e}}$ of up quarks, down
quarks and charged leptons, respectively. The $F$ terms of the
Higgs doublets plus their soft-breaking terms induce the scalar
potential
\begin{eqnarray}
\label{higgs} V_{Higgs}= m_{H_u}^2\, H_u^{\dagger} H_u +
m_{H_d}^2\, H_d^{\dagger} H_d + m_{u d}^2 \left(H_u \cdot H_d +
\mbox{h.c.}\right) + \mbox{D terms}
\end{eqnarray}
where the soft bilinear coupling $m_{u d}^2$, which can always be
made real by a rephasing of $H_{u,d}$, mixes the two doublets. The
mass splitting between the light and heavy physical Higgs doublets
depends on the strength of this mixing. In fact, the light Higgs
doublet, $h$, acquires a mass
\begin{eqnarray}
2\, m_h^2 = m_{H_u}^2 + m_{H_d}^2 - \sqrt{ \left(m_{H_u}^2 +
m_{H_d}^2\right)^2 +4 \left( m_{u d}^4 - m_{H_u}^2
m_{H_d}^2\right)}
\end{eqnarray}
which can be forced to lie right at the electroweak scale under
the admitted fine-tunings (\ref{fine}). Indeed, $h$ mimics the SM
Higgs doublet \cite{ad,followups} if
\begin{eqnarray}
m_{u d}^4 - m_{H_u}^2 m_{H_d}^2 \simeq {\rm TeV}^2\,
\left(m_{H_u}^2 + m_{H_d}^2\right)
\end{eqnarray}
where ${\rm TeV}$ stands for the Fermi scale. Consequently, when
the masses and mixings of the original Higgs doublets are
fine-tuned with an accuracy $\epsilon_h \sim {\rm
TeV}^2/M_{SUSY}^2$ a light tachyonic Higgs doublet emerges
automatically. The heavy Higgs doublet weighs ${\cal{O}}(M_{SUSY})$ and
its effect on the infrared (IR) dynamics is highly suppressed.

In the electroweak vacuum, $\langle h^0 \rangle \simeq m_{top}$,
all quarks and charged leptons acquire masses: ${\bf m_{u}} =
\sin\beta \langle h^0 \rangle {\bf Y_u}$, ${\bf m_{d, \ell}} =
\cos\beta \langle h^0 \rangle {\bf Y_{d, e}}$ where $\beta$ being the
Higgs mixing angle \cite{ad,followups}. The unitary rotations of
the superfields
\begin{eqnarray}
\label{ckmrot}
&&\widehat{Q} \rightarrow \left(\begin{array}{cc} V_{Q_U} & 0 \\
0 & V_{Q_D}\end{array}\right)\widehat{Q} \;,\; \widehat{L}
\rightarrow V_{L} \widehat{L}\;,\; \widehat{U}^c \rightarrow V_U\;
\widehat{U}^c \; ,\; \widehat{D}^c \rightarrow V_D\; \widehat{D}^c
\;, \; \widehat{E}^c \rightarrow V_E\; \widehat{E}^c
\end{eqnarray}
subject to the constraints
\begin{eqnarray}
\label{yukdiag} V_{Q_U}^{\bf T} {\bf Y_u} V_U = {\cal
Y}_u\;,\;\;V_{Q_D}^{\bf T} {\bf Y_d} V_D = {\cal Y}_d \;,\;\;
V_{L_E}^{\bf T} {\bf Y_e} V_E = {\cal Y}_e\;,
\end{eqnarray}
project all fermions into their physical bases via strictly
diagonal ${\cal Y}_{u,d,e}$. Under these transformations, the
neutral current vertices remain flavor-diagonal as in the gauge
basis whereas charged current vertices of quarks shuffle different
flavors via the CKM matrix $V_{CKM} = V_{Q_U}^{\dagger} V_{Q_D}$.
This very basis, the super-CKM basis, is highly useful for
analyzing the supersymmetric flavor mixings as additional 
sources with respect to  the standard flavor violation. 
Indeed, under (\ref{ckmrot}), the soft mass-squareds ${\bf M_{Q,\dots,E}^2}$ of
$\widetilde{Q},\dots,\widetilde{E}^c$ as well as their triscalar
couplings ${\bf Y_{u,d,e}^A}$ get dressed by the associated
unitary matrices. These dressings modify the gauge-basis flavor
structures significantly. For instance, the left-chirality squarks
$\widetilde{Q}_{up} \equiv \widetilde{U}_L$ and
$\widetilde{Q}_{down} \equiv \widetilde{D}_L$ acquire distinct
masses ${\bf M_{U_L}^2} = V_{Q_U}^{\bf T} {\bf M_{Q}^2}
V_{Q_U}^{\star}$ and ${\bf M_{D_L}^2} = V_{Q_D}^{\bf T} {\bf
M_{Q}^2} V_{Q_D}^{\star}$ even if ${\bf M_Q^2}$ is strictly
diagonal. Furthermore, couplings of neutral and charged components
of Higgs fields differ $e.g.$ $H_u^0$ ($H_u^+$) couples to
$\widetilde{U}_L$ $\widetilde{U}^{c}$ $\left(\widetilde{D}_L\,
\widetilde{U}^{c}\right)$ with $V_{Q_{U}}^{\bf T} {\bf Y_{u}^A}
V_{U} \equiv {\bf Y_u^A}$ $\left(V_{Q_{D}}^{\bf T} {\bf Y_{u}^A}
V_{U} \equiv {\bf \tilde{Y}_u^A}\right)$. Consequently,
corresponding to superpotential (\ref{superpot}), in super-CKM
basis, the most general holomorphic and $R$ parity conserving
operator structures parameterizing soft supersymmetry breaking in
the scalar fermion sector are given by
\begin{eqnarray}
\label{soft}
&&\widetilde{U}_L^{\bf T} {\bf M_{U_L}^2}
\widetilde{U}_L^{\star}\: + \widetilde{D}_L^{\bf T} {\bf
M_{D_L}^2} \widetilde{D}_L^{\star}\:+ \widetilde{U}^{c\, \dagger}
{\bf M_{U_R}^2} \widetilde{U}^c \:+\widetilde{D}^{c\, \dagger}
{\bf M_{D_R}^2} \widetilde{D}^c\:+ \widetilde{L}^{\bf T} {\bf
M_{L_L}^2} \widetilde{L}^{\star}\: +
\widetilde{E}^{c\, \dagger} {\bf M_{L_R}^2} \widetilde{E}^c\nonumber\\
&&+\left[ H_u^0 \widetilde{U}_L {\bf Y_u^A} \widetilde{U}^{c}
\:+{H}_d^0 \widetilde{D}_L {\bf Y_d^A} \widetilde{D}^{c}\: +
{H}_d^0 \widetilde{L}_L {\bf Y_e^A} \widetilde{E}^{c} +
\mbox{h.c.}\right]
\end{eqnarray}
where only the neutral Higgs couplings are displayed. The
soft-breaking masses ${\bf M_{U_L,\dots,L_R}^2}$ and ${\bf
Y_{u,d,e}^A}$ are $3\times 3$ matrices in flavor space. The mass
matrices are obviously hermitian. They conserve parity and
contribute to both CP and flavor violations via their
off--diagonal entries. On the other hand, the triscalar couplings
${\bf Y_{u,d,e}^A}$, like the Yukawas themselves, are general
parity-violating non-hermitian matrices. These flavor structures are
generic; neither their textures nor their correlations are known
{\it a priori}. They are completely independent of each other;
one's texture is independent of another's. The gauge-basis soft
mass-squareds ${\bf M_{Q,\dots, E}^2}$ break no symmetry but
supersymmetry; therefore, there is no smallness argument for any
of their entries due to the absence of chiral protection. On the
other hand, ${\bf Y_{u,d,e}^A}$ explicitly break the continuous
$R$ invariance of the theory hence their chiral protection.

Given that the mechanism which breaks the supersymmetry is not
known, use of the experimental bounds is the only way for
reconstructing the moduli and phases of ${\bf
M_{U_L,\dots,L_R}^2}$ and ${\bf Y_{u,d,e}^A}$. Indeed, if
supersymmetry survives down to the Fermi scale then these flavor
matrices induce sizeable modifications in rare processes (scaling
as $1/M_{SUSY}$ to appropriate power \cite{masiero}) as well as in
Higgs--fermion couplings (scaling as $|\mu|
|triscalars|/M_{SUSY}^2$ or $|\mu| |M_{gaugino}|/M_{SUSY}^2$
\cite{demir}). Consequently, if supersymmetry is not a weak-scale
invariance of Nature then constraints from the former fade away
whereas those from the latter remain intact. In the framework of
\cite{ad}, however, Higgs-fermion couplings, too, turn out to be
completely insensitive to supersymmetric flavor structures due to
the chiral protection of dimension-3 soft-breaking terms.
Irrespective of the scale of supersymmetry breaking, vacuum
stability arguments \cite{cd} impose rather strong constraints on
the off-diagonal entries of triscalar couplings; however, such
bounds are automatically satisfied in the framework of \cite{ad}.
In light of these observations one concludes that flavor mixings
in the sfermion sector are rather generic and remain unconstrained
by phenomenological bounds.

Having summarized the status of flavor violation in low-scale
supersymmetry, we now discuss implications of flavor mixings in
${\bf M_{U_L,\dots,L_R}^2}$ for sparticle spectrum when
fine-tunings ${\cal{O}}\left(\epsilon_h\right)$ are built-in
properties of Nature. It is convenient to analyze first a rather 
plain flavor structure by specializing to one of the mass-squareds $e.g.$
${\bf M_{U_L}^2}$:
\begin{eqnarray}
\label{qmass} {\bf M_{U_L}^2} = \left(\begin{array}{ccc} m_{{1}}^2
& 0 &
0\\\\
0 & m_{{2}}^2 & m_{{2 3}}^2\\\\ 0 & m_{{2 3}}^{2\, \star} &
m_{{3}}^2\end{array}\right)
\end{eqnarray}
with two texture zeroes and  $m_{{1,2,3}}^2 \sim \left|m_{{2
3}}^2\right|\sim M_{SUSY}^2$. That the second and third
generations exhibit such a strong mixing implies that the mass
\begin{eqnarray}
2\, m_{\widetilde{u}_L}^2 = m_{{2}}^2 + m_{{3}}^2 - \sqrt{
\left(m_{{2}}^2 + m_{{3}}^2\right)^2 +4 \left( {|m_{{2 3}}^2|}^2 -
m_{{2}}^2 m_{{3}}^2\right)}
\end{eqnarray}
of the lightest flavor eigenstate, with $\gamma_{u_L} =
\mbox{Arg}[m_{{2 3}}^2]$,
\begin{eqnarray}
\widetilde{u}_L\equiv \cos\theta_{u_L}\, \widetilde{U}_{L_2} +
\sin\theta_{u_L}\, e^{- i \gamma_{u_L}}\, \widetilde{U}_{L_3}
\end{eqnarray}
can be fine-tuned to lie right at the weak scale if
\begin{eqnarray}
m_{2}^2 m_{3}^2 - |m_{{2 3}}^2|^2 \simeq {\rm TeV}^2\, \left(
m_{2}^2 + m_{3}^2\right)
\end{eqnarray}
holds. The amount of fine-tuning involved here is of the same size
as the one needed for generating a light Higgs doublet. The
weak-scale effective theory is obtained by the replacements
\begin{eqnarray}
\widetilde{U}_{L_2} \rightarrow \cos \theta_{u_L}\,
\widetilde{u}_L\; , \;\; \widetilde{U}_{L_3} \rightarrow \sin
\theta_{u_L}\, e^{i \gamma_{u_L}}\, \widetilde{u}_L
\end{eqnarray}
in the lagrangian. Clearly, dressing of the interaction vertices
by $\cos \theta_{u_L}$ or $\sin \theta_{u_L}$ signals the
decoupling of heavy squark up squark from the light spectrum.

In general, ${\bf M_{U_L}^2},\dots, {\bf M_{L_R}^2}$ are
independent matrices, and thus, none, part or all of them can
exhibit a structure similar to (\ref{qmass}). In case each of them
accommodates a strong mixing between any two flavors then the
weak-scale effective theory consists of seven distinct sfermion
states within the reach of LHC. Of course, all these light
sfermions are only flavor eigenstates; whether they are physical
fields or not depend on if triscalar couplings and $F$ terms feed
further mixings. To see this point in detail, suppose that
intergenerational mixings are all small except that ${\bf
M_{U_L}^2}$ and ${\bf M_{U_R}^2}$ possess sizeable (2,3) and (1,2)
entries, respectively. Then, the light sparticle spectrum consists
of two up squarks $\widetilde{u}_L$ and $\widetilde{u}_R$ of
opposite chirality and a single Higgs doublet $h$ in excess of
gauginos and Higgsinos. An interesting aspect of this spectrum is
that squarks develop left-right mixings via the triscalar
couplings in (\ref{soft}) and $F$ term contributions. Indeed, in
$\left( \widetilde{u}_{L}, \widetilde{u}_R\right)$ basis the mass-squared
matrix of up squarks takes the form
\begin{eqnarray}
\label{umass} {\bf M^2_{\widetilde{u}}} = \left( \begin{array}{cc}
m_{\widetilde{u}_L}^2 +\Delta_L\, \langle h^0
\rangle^2 & \sin \beta\, \langle h^0 \rangle\, m_{LR} \\
\sin \beta\, \langle h^0 \rangle\, m_{LR}^{\star} &
m_{\widetilde{u}_R}^2 +\Delta_{R}\, \langle h^0 \rangle^2
\end{array}\right)
\end{eqnarray}
where
\begin{eqnarray}
\Delta_L &=& \cos^2\beta \left|\left({\cal{Y}}_u\right)_{2 2} \cos
\theta_{u_L} + \left({\cal{Y}}_u\right)_{3 3} \sin \theta_{u_L}
e^{i \gamma_{u_L}}\right|^2 + \frac{1}{4} \cos 2\beta \left( g_2^2
- \frac{1}{3} g_1^2\right)\nonumber\\
\Delta_R &=& \cos^2\beta \left|\left({\cal{Y}}_u\right)_{1 1} \cos
\theta_{u_R} + \left({\cal{Y}}_u\right)_{2 2} \sin \theta_{u_R}
e^{i \gamma_{u_R}}\right|^2 + \frac{1}{3} \cos 2\beta  g_1^2
\end{eqnarray}
are generated by sfermion $F$ and $D$ terms quadratic in 
$\widetilde{u}_{L,R}$, and
\begin{eqnarray}
m_{LR} &=& \left({\bf Y_u^A}\right)_{3 1} \sin\theta_{u_L}
\cos\theta_{u_R} e^{i \gamma_{u_L}} + \left({\bf Y_u^A}\right)_{3
2} \sin\theta_{u_L} \sin\theta_{u_R} e^{i(\gamma_{u_L}+\gamma_{u_R})}\nonumber\\
&+& \left({\bf Y_u^A}\right)_{2 1} \cos\theta_{u_L}
\cos\theta_{u_R} + \left({\bf Y_u^A} - \mu^{\star} \cot\beta\,
{\cal{Y}}_u\right)_{2 2} \cos\theta_{u_L} \sin\theta_{u_R} e^{i
\gamma_{u_R}}
\end{eqnarray}
is induced by the $F$ terms of Higgs fields and triscalar
couplings in (\ref{soft}). This quantity receives contributions
from all relevant entries of ${\bf Y_u^A}$ not just from its (2,2)
corner, and it comprises phases of both large and small scales of
the model. The physical up squark states are achieved by
diagonalizing (\ref{umass}); its mixing matrix will involve
$\mbox{Arg}[m_{LR}]$ to contribute to CP violation in processes
proceeding with $\widetilde{u}_{L,R}$ mediation.

Going to flavor-diagonal basis directly modifies the supergauge
vertices even if squarks do not exhibit any LR mixing. Indeed, the
light squarks $\widetilde{u}_{L,R}$ above interact with gluino and
quarks via
\begin{eqnarray}
\label{int} \cos\theta_{u_L} \widetilde{u}_L
\overline{\widetilde{g}^a_R} {\lambda^a} c_L + \sin\theta_{u_L}
e^{- i \gamma_{u_L}} \widetilde{u}_L \overline{\widetilde{g}^a_R}
{\lambda^a} t_L - \cos\theta_{u_R} \widetilde{u}_R
\overline{\widetilde{g}^a_L} {\lambda^a} u_R - \sin\theta_{u_R}
e^{-i \gamma_{u_R}} \widetilde{u}_R \overline{\widetilde{g}^a_L}
{\lambda^a} c_R
\end{eqnarray}
in units of $- g_{s}/\sqrt{2}$. It is highly remarkable that a
given light squark couples to distinct quark fields: a crucial
property for flavor-changing neutral current (FCNC) transitions.
In general, gaugino-fermion-sfermion couplings of this form
generically exist for down quark and lepton sectors, too, provided
that their mass-squared matrices exhibit two texture zeroes like
(\ref{qmass}).

The results derived above are not special to the mixing pattern in
(\ref{qmass}). Indeed, mass-squared matrices having a single texture zero
\begin{eqnarray}
\label{1zero}
\left(\begin{array}{ccc} \# & 0 & \# \\
0 & \# & \# \\
\# & \# & \# \end{array}\right)\; \; , \;\;\; \left(\begin{array}{ccc} \# & \# & 0 \\
\# & \# & \# \\
0 & \# & \# \end{array}\right) \; \; , \;\;\; \left(\begin{array}{ccc} \# & \# & \# \\
\# & \# & 0 \\
\# & 0 & \# \end{array}\right)
\end{eqnarray}
with all $\#$'s being ${\cal{O}}(M_{SUSY})$, also possess one
small eigenvalue ${\cal{O}}({\rm TeV})$ and two large ones
${\cal{O}}(M_{SUSY})$ provided that fine-tunings
${\cal{O}}(\epsilon_h)$ are admitted. For such flavor structures,
all three generations of sfermions of a given chirality possess a
light component, and that light sfermion interacts with gauginos
and all three generations of fermions in a way generalizing
(\ref{int}). Furthermore, if at least one of $\left({\bf
M_{U_L}^2}, {\bf M_{U_R}^2}\right)$, $\left({\bf M_{D_L}^2}, {\bf
M_{D_R}^2}\right)$, $\left({\bf M_{L_L}^2}, {\bf
M_{L_R}^2}\right)$ exhibit flavor mixings as in (\ref{1zero}) then
the resulting opposite-chirality light sfermions develop
left-right mixings such that $m_{LR}$ now involves all entries of
triscalar couplings.

In addtion to (\ref{qmass}) and (\ref{1zero}), the sfermion mass-squared  matrices
can assume a democratic texture as well
\begin{eqnarray}
\label{0zero}
\left(\begin{array}{ccc} \# & \# & \# \\
\# & \# & \# \\
\# & \# & \# \end{array}\right)
\end{eqnarray}
which accommodates a heavy state weighing ${\cal{O}}(M_{SUSY})$
and two light states with masses ${\cal{O}}({\rm TeV})$. If all of
${\bf M_{U_L}^2},\dots, {\bf M_{L_R}^2}$ exhibit democratic flavor
mixings then light spectrum contains a total of fourteen sfermion
states. In case, for instance, ${\bf M_{U_L}^2}$ and ${\bf
M_{U_R}^2}$ are of the form (\ref{0zero}) then there will be four
light up squark states and they will experience both intra-- and
inter--generational mixings in the left-right block in a way
involving all entries of ${\bf Y_u^A}$. Furthermore, each of these
light up squarks will interact with all three up quarks via
supergauge vertices.

Having derived the light sparticle  spectrum arising from
${\cal{O}}(M_{SUSY})$ flavor mixings,  we now discuss their
phenomenological implications:
\begin{itemize}
\item Even if there exists a single light sfermion flavor,
fermion-sfermion-gaugino vertices necessarily lead to FCNC
transitions. This is clear from (\ref{int}) in which a light
sfermion develops interactions with fermions of varying flavor.
Consider, for instance, the last two terms of (\ref{int}) which
involve couplings of right-handed $u$ and $c$ quarks to a single
up type squark. Obviously, these vertices generate supersymmetric
contributions to rare processes involving $u_R$--$c_R$ transition.
For instance, the effective Hamiltonian for
$D^0$--$\overline{D^0}$ mixing receives the contribution
\begin{eqnarray}
\label{ccbar} \Delta H_{eff} = \alpha_s^2\, \cos^2\theta_{u_R}\,
\sin^2 \theta_{u_R}\, e^{i 2 \gamma_{u_R}}\,
\frac{1}{\left|M_{\widetilde{g}}\right|^2} \left(
\frac{x+9}{18(1-x)^2} + \frac{3 x +2}{9 (1-x)^3} \ln x\right)
\left(\overline{c_R} \gamma_{\mu} u_R\right)^2
\end{eqnarray}
which contributes to both the lifetime and CP impurity of the
neutral $D$ meson. This operator can lead to significant effects
depending on the sizes of
$x=m_{\widetilde{u}_R}^2/\left|M_{\widetilde{g}}\right|^2$ and the
gluino mass so that the present bounds can put stringent limits on
$\sin 2 \theta_{u_R}$ \cite{masiero}. Obviously, in case the
associated mass-squared matrices admit light states the structure
above repeats for other meson mixings as well. For example, if
${\bf M_{D_L}^2}$ exhibits a texture like (\ref{qmass}) then
$B_s^0$--$\overline{B_s^0}$ mixing receives a contribution similar
to (\ref{ccbar}) with the replacements $\theta_{u_R} \rightarrow
\theta_{d_L}$, $\gamma_{u_R} \rightarrow \gamma_{d_L}$,
$m_{\widetilde{u}_R} \rightarrow m_{\widetilde{d}_L}$ and
$\overline{c_R} \gamma_{\mu} u_R \rightarrow \overline{b_L}
\gamma_{\mu} s_L$. In addition to mixings of mesons, their decays
are also influenced by light sparticle spectrum since, for
instance, the first two terms of (\ref{int}) give rise to the rare
decays $t_L \rightarrow c_L (\gamma, g)$.

Depending on how crowded the light spectrum is, rare processes
could be mediated by various sfermions. For instance, when there
are two squarks experiencing LR mixings as in (\ref{umass}) then
meson mixings receive contributions from $(V-A)\otimes(V-A)$,
$(V+A)\otimes(V+A)$ and $(V-A)\otimes(V+A)$ type $\Delta F = 2$
operators. Furthermore, the flavor textures (\ref{1zero}) and
(\ref{0zero}) enable light sfermions to couple to all three
fermions of given electric charge and thus they contribute to
mixings and decays of more than one meson such that the sfermion
line can flip both chirality and flavor \cite{masiero2}.

\item The very existence of light sfermion states necessitates
radiative corrections to Higgs boson mass beyond those in the SM.
This statement is valid even if there exists a single sfermion
field at the ${\rm TeV}$ scale. For exemplifying this point,
consider a light $\widetilde{u}_L$ springing from (\ref{qmass}).
At one loop this state shifts the Higgs boson mass by an amount
\begin{eqnarray}
\label{dmh} \delta m_h^2 = \frac{3}{8 \pi^2}\, \Delta_L \left\{
\left(m_{\widetilde{u}}^2 + 2 \Delta_L \langle h^0\rangle^2\right)
\ln \frac{m_{\widetilde{u}}^2}{Q^2} - m_{\widetilde{u}}^2\right\}
\end{eqnarray}
at the renormalization scale $Q$, in the ${\overline{{\rm MS}}}$
scheme, with physical squark mass $m_{\widetilde{u}}^2 =
m_{\widetilde{u}_L}^2 + \Delta_L \langle h^0\rangle^2$. It is
straightforward to generalize (\ref{dmh}) to cases with several
sfermions of varying flavor and chirality. Clearly, unlike the
conventional MSSM \cite{higgs}, the Higgs sector does not violate
CP invariance.

\item The electric dipole moments (EDM) are no exception; they are
induced already at one-loop level in the presence of light
sfermions. Indeed, the first and last terms of (\ref{int}), for
instance, induce a finite EDM for charm quark via gluino-squark
exchange:
\begin{eqnarray}
\frac{{\cal{D}}_{c}}{e} &=& - \frac{4 \alpha_s}{9 \pi}\, \frac{
\left|m_{LR}\right| \langle h^0
\rangle}{\left|M_{\widetilde{g}}\right|^3}\, \cos \theta_{u_L}\,
\sin \theta_{u_R}\, \sin \beta\, \sin \left(\gamma_{LR} +
\gamma_{u_R} - \gamma_{\widetilde{g}}\right)\nonumber\\ &\times&
\frac{x_1 x_2}{x_1-x_2} \left( f\left(x_1\right) -
f\left(x_2\right)\right)
\end{eqnarray}
where $m_{\widetilde{u}_{i}}^2$, with $i=1,2$, are the eigenvalues
of (\ref{umass}), $x_i =
m_{\widetilde{u}_i}^2/\left|M_{\widetilde{g}}\right|^2$ and
$f(x)=\left(x/2 (1-x)^3\right)$ $\left(1-x^2 + 2 x \ln x\right)$.
The EDMs, though genuinely flavor-diagonal, gain a direct
sensitivity to the flavor mixings in triscalar couplings via its
dependence on the modulus and phase of $m_{LR}$. Clearly, as the
light sfermions populate in flavor and chirality, the EDMs of
fermions are contributed by a variety of diagrams \cite{edm}.

\item The question of if and how gauge couplings unify depends
on the light spectrum and on the scale at which heavy spectrum 
lies. For the model under concern, the light spectrum is 
composed of SM fermions (which come in complete SU(5) irreps),
the gauginos and Higgsinos, a light Higgs doublet and a set 
of light sfermions. Therefore, the one-loop beta function 
coefficients for SU(3)$_c$, SU(2)$_L$ and U(1)$_Y$ are given, 
respectively, by $b_3= -5 +\Delta_3/3$, $b_2= -7/6 +
\Delta_2/3$ and $b_1= 9/2 + \Delta_1/3$ where $\Delta_i$ 
represent the contributions of light sfermions 
(they all vanish in the `minimal' split supersymmetry 
\cite{ad,followups}). A straightforward calculation gives
$\left(\Delta_3, \Delta_2, \Delta_1\right)=(1/2, 1/2, 1/20)$,
$(1/2, 0, 4/5)$, $(1/2, 0, 1/5)$, $(0, 1/2, 3/20)$, $(0, 0, 3/5)$
for a light $\widetilde{u}_L$, $\widetilde{u}_R$, $\widetilde{d}_R$,
$\widetilde{e}_L$ and $\widetilde{e}_R$, respectively. Consequently,
each gauge coupling runs with a different slope depeding on the type
of the light sfermion field. For recovering complete gauge coupling 
unification, the MSSM unification scale must lie above the 
scale of heavy scalars between which the running is precisely that 
of the MSSM \cite{ad,followups}.

\item The light spectrum will leave observable signatures at the
LHC; however, unlike the predictions of flavor-conserving split
supersymmetry \cite{ad}, gluino does not need to be a long-lived
fermion; moreover, LHC does not need to be a gluino factory; such
statements are in variant with the mass spectra of inos and
sfermions. The LHC will produce sfermions as well though they
differ from the conventional MSSM spectrum in population and
couplings. It is interesting to compare predictions of flavor-blind
and flavor-violating split-supersymmetric models in connection 
with \cite{cosmic}. Indeed, in this work it was found that long-lived
gluinos, produced in distant galactic nuclei, induce cosmic ray 
showers with rates detectable in upcoming Pierre Auger Observatory. 
Therefore, in case such signals are experimentally confirmed, 
it will be possible to conclude that the mechanism that breaks 
supersymmetry is flavor-blind. In the opposite case, we will 
have to live with light sparticles which can facilitate a 
number of processes listed above.

\end{itemize}
By explicit examples ranging from rare processes to collider
expectations, we have briefly discussed certain phenomenological
consequences of supersymmetric models endowed with ultra heavy
scalars and nontrivial flavor mixings when Nature admits
fine-tunings in (\ref{fine}). Our findings show that, flavor
mixings give rise to light sfermion states whose contributions to
various observables can be bounded even with the present level of
experimental exploration. The ${\rm TeV}$ scale effective theory
has interesting differences from the conventional MSSM in terms of
the couplings, masses and CP properties of the particles. It is
after a global analysis of the existing bounds on FCNCs, Higgs
mass, EDMs, $\dots$ that one can determine the likelihood ranges
of various model parameters.

In conclusion, if the supersymmetric spectrum is to be completely
split, that is, if gauginos, Higgsinos and a Higgs doublet are to be 
the only sparticles to weigh at the Fermi scale then one must prevent
flavor mixings among the sfermions at $M_{SUSY}$ in addition to
implementing chiral symmetries for protecting dimension-3 soft
terms. Recently, the work of Arkani-Hamed and Dimopoulos 
has been rectified by constructing explicit models 
\cite{adgr} in the spirit of \cite{gm}. It has there been shown that, 
spurion fields of the form $\widehat{X} = 1 + \theta^4 M_{SUSY}^2$ 
break supersymmetry but preserve the $R$ invariance in accord with the requirements of splitting the
sparticle spectrum. At the renormalizable level, such spurion
fields couple to the visible sector fields either as $\int d^4
\theta \widehat{Q}^{\dagger}_i \widehat{X}_{i j} \widehat{Q}_j$ or
as  $\int d^4 \theta \widehat{H}_u \widehat{X}_{u d}
\widehat{H}_d$ where $\widehat{X}_{i j, u d}$ are dimensionless
superfields. These interactions generate the soft-mass squareds
$\left({\bf M_Q^2}\right)_{i j}$ and $m_{u d}^2$, all being
naturally ${\cal{O}}(M_{SUSY})^2$. The textures of the sfermion
mass-squareds is determined by $D$ components of $\widehat{X}_{i
j}$; in case spurions with $i \neq j$ are sufficiently suppressed
compared to those with $i= j$ then weak-scale effective theory is
precisely the one put forward in \cite{ad}. However, in the
absence of a symmetry principle that can impose such a hierarchy
it will not be possible to clean up the Fermi scale from
sfermions. A candidate symmetry would be the invariance of entire
sfermion sector (excluding the triscalar couplings) 
under unitary rotations of the form $\widetilde{f}
\rightarrow U\, \widetilde{f}$ with $U U^{\dagger} = {\bf 1}$.
Indeed, such an invariance  would be operative if and only
if the gauge-basis soft mass-squareds are proportional to the
identity matrix. Putting differently, the sfermion mass-squareds
must remain unchanged as one switches from gauge basis to
super-CKM basis. Clearly, this symmetry does not need to be exact;
it can be an approximate symmetry provided that off-diagonal entries
of the sfermion mass-squared matrices are not comparable with those 
at the diagonal.

\vspace{1cm} The author greatfully acknowledges enlightening
e-mail exchanges with Nima Arkani-Hamed about finely-tuned Higgs
proposed in \cite{ad}. This work was supported by the GEBIP grant
of Turkish Academy of Sciences.

\end{document}